\documentclass[fleqn,usenatbib]{mnras} 
\usepackage{graphicx} 
\usepackage{amsmath} 
\usepackage{amssymb} 
\usepackage{color}
\usepackage[T1]{fontenc} 
\usepackage{ae,aecompl} 
\usepackage[dvipsnames]{xcolor}
\renewcommand{\vec}[1]{\bmath{#1}}
\newcommand{\altered}[1]{{#1}}
\newcommand{\alteredb}[1]{{#1}}

\newcommand{\epsnorm}{\overline{\epsilon^{\rm s}}^2}

\title{Intrinsic alignment contamination to CMB lensing--galaxy weak lensing correlations from tidal torquing}
\author[Patricia~Larsen and Anthony Challinor]
{Patricia~Larsen,$^1$\thanks{prl37@cam.ac.uk} and Anthony~Challinor$^{1,2}$\thanks{a.d.challinor@ast.cam.ac.uk}\\
$^1$Institute of Astronomy and Kavli Institute for Cosmology Cambridge, Madingley Road, Cambridge, CB3 0HA, UK\\
$^2$DAMTP, Centre
for Mathematical Sciences, Wilberforce Road, Cambridge CB3 0WA, UK}

\begin{document}
\label{firstpage}
\pagerange{\pageref{firstpage}--\pageref{lastpage}}
\maketitle

\begin{abstract}
Correlations of galaxy ellipticities with large-scale structure, due to galactic tidal
interactions, provide a potentially 
significant contaminant to measurements of cosmic shear. However, these intrinsic alignments are still poorly understood for galaxies at the redshifts typically used in cosmic shear analyses. For spiral galaxies, it is thought that tidal torquing is significant in determining alignments resulting in zero correlation between the intrinsic ellipticity and the gravitational potential in linear theory. Here, we calculate the leading-order correction to this result in the tidal-torque model from non-linear evolution, using second-order perturbation theory, and relate this to the contamination from intrinsic alignments to the recently-measured cross-correlation between galaxy ellipticities and the CMB lensing potential. On the scales relevant for CMB lensing observations, the squeezed limit of the gravitational bispectrum dominates the correlation. Physically, the 
large-scale mode that sources CMB lensing modulates the small-scale power and hence the intrinsic ellipticity, due to non-linear evolution.  We find that the angular cross-correlation from tidal torquing has a very similar scale dependence as in the linear alignment model, believed to be appropriate for elliptical galaxies. The amplitude of the cross-correlation is predicted to depend strongly on the formation redshift, being smaller for galaxies that formed at higher redshift when the bispectrum of the gravitational potential was smaller. Finally, we make simple forecasts for constraints on intrinsic alignments from the correlation of forthcoming cosmic shear measurements with current CMB lensing measurements. We note that cosmic variance can be significantly reduced in measurements of the difference in the intrinsic alignments for elliptical and spiral galaxies if these types can be separated (e.g., using colour).
\end{abstract}

\section{INTRODUCTION}
Gravitational weak lensing by large-scale structure has emerged in recent years as a powerful tool for studying structure formation and the 
expansion history of the universe. It coherently distorts images of background galaxies, magnifying their angular size and shearing their 
intrinsic projected shapes, in a manner proportional to the integrated total mass density along the line of sight back to the source galaxy. The current generation of imaging surveys, including DES\footnote{http://www.darkenergysurvey.org/}, KiDS\footnote{http://kids.strw.leidenuniv.nl/} and HSC\footnote{http://www.naoj.org/Projects/HSC/HSCProject.html}, will significantly improve the statistical power of cosmic shear measurements, and will be followed by much larger surveys (\textit{EUCLID}\footnote{http://www.cosmos.esa.int/web/euclid} and LSST\footnote{http://www.lsst.org/lsst/}) in the next decade.


Gravitational lensing also affects the temperature anisotropies and polarization of the cosmic microwave background (CMB); see~\citet{lewischallinor06} for a review. The lensing deflections can be reconstructed from quadratic combinations of the CMB data~\citep{Zaldarriaga:1998te,Hu:2001tn,Hu:2001kj}, and the power spectrum of such reconstructions has now been demonstrated to be a powerful cosmological probe~\citep{2011PhRvL.107b1302S,2014A&A...571A..17P,2015arXiv150201591P}. CMB lensing is highly complementary to galaxy lensing, providing measurements of the matter distribution at higher redshifts and closer to the linear regime. The combination of the two lensing measurements is particularly powerful~\citep{kitching14,vallinotto12,vallinotto13}. Their cross-correlation probes structure at intermediate redshifts ($z\approx 1$), and should be much more immune to systematic effects than either auto-correlation. The CMB lensing--galaxy weak lensing correlation has recently been detected by~\citet{hand13}, using a CMB lens reconstruction from ACT data and galaxy lensing maps from the CFHT Stripe 82 Survey, by~\citet{Liu:2015xfa} using reconstructions from \textit{Planck} data and galaxy lensing from CFHTLenS, and by~\citet{Harnois-Deraps:2016huu} using the \textit{Planck} lensing reconstruction and galaxy lensing from CFHTLenS and RCSLenS.

Correlations of the intrinsic (i.e., before lensing) ellipticities of galaxies, termed intrinsic alignments, are known to contaminate galaxy lensing surveys (see \citealt{troxelishakreview} for a recent review and~\citealt{2015arXiv150405456J,2015arXiv150405546K,2015arXiv150405465K}, which form a dedicated topical volume on the subject). Intrinsic alignments are generally thought to arise from tidal fields acting at the time of galaxy formation, although this is not certain (see \citealt{camelio15} and references therein). Nearby galaxies that formed in the same tidal field will have correlations between their intrinsic ellipticities, contributing what is known as the II term to the observed shear correlations. A further contribution, known as the GI term, arises from the tidal field in which a foreground galaxy forms also contributing to the gravitational shearing of a background galaxy~\citep{hirataseljak04}.

\altered{The CMB lensing--galaxy lensing cross-correlation is contaminated solely by a term equivalent to the GI term.} This signal, as it applies to the \citet{hand13} measurement, has been estimated analytically by \citet{hall} and \citet{iacross2} in the linear alignment model, giving a 15\% reduction in the observable correlation. In this model, the shape of a galaxy is assumed to be determined by the shape of its host halo, with the latter assumed to be linearly related to the gravitational tidal field through tidal stretching at the epoch of formation~\citep{catelan}. The linear-alignment model is motivated for elliptical galaxies, and has been successful in modelling the observed number density--shape correlations of Luminous Red Galaxies (see e.g.,~\citealt{Singh:2014kla} for recent work). 

A large proportion of galaxies at the redshifts associated with galaxy surveys are, however, dim, blue galaxies that are often spirals. \altered{In this paper we use the terms blue and spiral galaxies interchangeably for simplicity, despite noting that blue ellipticals exist.} 
For spiral galaxies, the intrinsic ellipticity is determined by the disc orientation and so the angular momentum direction. Tidal-torque theory describes the generation of halo angular momentum via the torque exerted on a forming halo for which the inertia tensor is misaligned with the tidal tensor (see e.g.,~\citealt{1984ApJ...286...38W}). The intrinsic ellipticity is quadratic in the components of the angular momentum in the plane of the sky, leading to an average ellipticity for a given tidal field that is quadratic in the tidal field~\citep{mwk}. 
\altered{The observed autocorrelations of such galaxies have shown hints of such a quadratic dependence on the tidal field at low redshifts~\citep{Lee:2007tq}, although the significance of these results has been disputed~\citep{andrae11}, and no direct detection has been made at redshifts around $z\approx 0.6$~\citep{2011MNRAS.410..844M}.}

In this \emph{quadratic alignment model} for spiral galaxies, the GI correlation vanishes in linear theory for which the gravitational potential is Gaussian~\citep{huizhang,hirataseljak04}.\footnote{In the more sophisticated model of~\citet{cnpt}, the average ellipticity is quadratic in the \emph{normalised} tidal field, but this does not change the result that the GI term vanishes in linear theory~\citep{huizhang}.} By modeling the fraction of elliptical galaxies in the CFHT Stripe 82 lensing maps, ~\citet{Chisari:2015bna} estimate a lower contamination of around 9\% from GI correlations to the CMB lensing--galaxy lensing measurement of~\citet{hand13}, assuming no GI contribution from spiral galaxies.
However, non-linear evolution generates non-Gaussianity in the potential, and so is expected to source some non-zero correlations between the intrinsic ellipticity and gravitational potential even for spiral galaxies. The size of this non-linear contribution to the number density--ellipticity correlation was estimated in~\citet{huizhang}. Very recently, correlations between the shapes of disc-like galaxies and overdensities of elliptical galaxies in hydrodynamical simulations have been reported~\citep{Chisari:2015qga, illustrus}. \altered{However, these studies differ in the nature of disc orientations. In the Horizon-AGN simulation,~\citet{Chisari:2015qga} find that the discs are preferentially tangentially oriented around overdensities, while in the Illustrus simulation,~\citet{illustrus} find a radial alignment.} 

Our goal in this paper is to estimate the non-linear GI contribution to CMB lensing--galaxy lensing measurements in the quadratic alignment model, and to assess whether this can be a significant contaminant. The paper is organised as follows. Section~\ref{sec:gravlens} briefly reviews gravitational lensing in the cosmological context, while Sec.~\ref{sec:linear} summarizes the intrinsic alignments in the linear alignment model, following closely \citet{hall}. Our main results are in Sec.\ \ref{sec:quad}, where we calculate the CMB lensing--ellipticity correlation in the quadratic alignment model, investigating the leading-order non-Gaussian contribution. 
Section~\ref{sec:quad} also includes detection forecasts for intrinsic alignments from CMB lensing--galaxy lensing measurements with forthcoming cosmic shear surveys.
Finally, we conclude in Sec.\ \ref{sec:summary}.

\section{GRAVITATIONAL LENSING FORMALISM}
\label{sec:gravlens}

Gravitational lensing alters the path of light rays, changing the direction from which they appear to be emitted. The associated change in angular coordinates in the plane of the sky is given by
\begin{equation}
\left( \begin{array}{c} \Delta x_{\mathrm{s}} \\ \Delta y_{\mathrm{s}} \end{array} \right) = \left( \begin{array}{cc}  1-\kappa -\gamma_1  & -\gamma_2 \\ -\gamma_2 & 1-\kappa+\gamma_1 \end{array} \right) \left( \begin{array}{c} \Delta x \\ \Delta y \end{array} \right),
\end{equation}
where the subscript s refers to coordinates in the source plane, and the changes on the right are in the observational coordinates. The shear terms $\gamma_1$ and $\gamma_2$ describe anisotropic, area-preserving distortions, while the convergence term $\kappa$ describes isotropic changes in angular size.
We take the components of the shear relative to unit vectors along the $\theta$ and $\phi$ directions of a spherical coordinate system. The complex shear $\gamma \equiv \gamma_1 + i \gamma_2$ is a spin $+2$ quantity, and, for sources at comoving distance $\chi$ in a spatially-flat universe, the shear along the line-of-sight $\hat{\vec{n}}$ is (in the Born approximation)
\begin{equation}
\gamma(\hat{\vec{n}};\chi) = \int_0^{\chi} d\chi' \, \frac{\chi-\chi'}{\chi \chi'} \eth^2 \Phi(\chi'\hat{\vec{n}}, \chi') \, .
\end{equation}
Here, $\Phi$ is the Newtonian potential at position $\chi' \hat{\vec{n}}$ at conformal lookback time $\chi'$, and $\eth$ is the spin-raising operator~(e.g.,~\citealt{goldberg1967}). For sources with a redshift distribution $f(\chi)$, the average shear along $\hat{\vec{n}}$ is
\begin{equation}
\gamma(\hat{\vec{n}}) = \int d\chi\, f(\chi) \gamma(\hat{\vec{n}};\chi)\, .
\end{equation}
The complex shear can be expanded in spherical multipoles as
\begin{equation}
\gamma(\hat{\vec{n}}) = \sum_{lm} \gamma_{lm} {}_2 Y_{lm}(\hat{\vec{n}}) \, ,
\end{equation}
where ${}_2 Y_{lm} = \sqrt{(l-2)!/(l+2)!} \eth^2 Y_{lm}$ are the spin-weight $2$ spherical harmonics. The shear has only $E$-modes, so that $\gamma_{lm}^\ast = (-1)^m \gamma_{l\, -m}$.

The shear can be estimated from the observed shape of galaxies.
For weak lensing, the observed ellipticity\footnote{For an elliptical source with minor-to-major axis ratio $r$, we define the ellipticity such that $|\epsilon| = (1-r)/(1+r)$.} is the sum of the intrinsic ellipticity $\epsilon^{\rm s}$ and the gravitational shear
\begin{equation}
\epsilon \approx \epsilon^{\rm s} + \gamma \, .
\end{equation}
In cosmological weak lensing, the gravitational shear is typically around 1\,\% or less, much less than the intrinsic ellipticity that has r.m.s.\ dispersion 10--20\,\%, so that a statistical treatment based on the correlation of the observed shapes of many galaxies is required.

For lensing of the CMB, the source redshift distribution is effectively a delta function at the time of last scattering, $f(\chi) = \delta(\chi-\chi_*)$. Writing the shear in terms of the CMB \emph{lensing potential} as $\gamma = -\eth^2 \phi/2$, we have
\begin{equation}
\phi(\hat{\vec{n}}) = -2 \int_{0}^{\chi_{*}} d\chi \, \frac{\chi_{*}-\chi}{\chi \chi_{*}}  \Phi(\chi \hat{\vec{n}},\chi) \, .
\label{eqn:cmblens}
\end{equation}
Expanding $\phi(\hat{\vec{n}})$ in spherical harmonics, $\phi(\hat{\vec{n}}) = \sum_{lm} \phi_{lm} Y_{lm}(\hat{\vec{n}})$, the cross-correlation with the galaxy shear is
\begin{equation}
\langle \gamma_{lm} \phi_{l'm'}^\ast \rangle = C_l^{\gamma \phi} \delta_{ll'} \delta_{mm'} \, ,
\end{equation}
where, in the Limber approximation,
\begin{equation}
C_l^{\gamma\phi} = -2 \sqrt{\frac{(l+2)!}{(l-2)!}} \int_0^{\chi_*} d\chi\, \frac{W(\chi)}{\chi^2} \frac{\chi_*-\chi}{\chi \chi_*} P_{\Phi}(l/\chi;\chi) \, ,
\label{eq:gammaphi}
\end{equation}
where
\begin{equation}
W(\chi) = \int_{\chi}^{\chi_*} d\chi' \, f(\chi') \frac{\chi'-\chi}{\chi \chi'} \, .
\end{equation}
Here, $P_{\Phi}(k,\chi)$ is the dimensional power spectrum of the gravitational potential at conformal lookback time $\chi$.
Measurements of the cross-correlation of the observed galaxy ellipticities with the CMB lensing potential give instead, on average,
\begin{equation}
C_l^{\epsilon \phi}  = C_l^{\gamma \phi}+ C_l^{\epsilon^{\rm s} \phi} \, ,
\end{equation}
where the second term arises from intrinsic alignments. It is analogous to the $GI$ term of \citet{hirataseljak04}, and will contaminate any measurements of the gravitational shear. Note that, generally, the observed ellipticity has both $E$- and $B$-modes, because of $\epsilon^{\rm s}$, but the $B$-modes do not contribute to the cross-correlation with $\phi$ due to parity. The following sections aim to calculate $C_l^{\epsilon^{\rm s} \phi}$ in the linear and quadratic alignment models.

\section{(NON-)LINEAR ALIGNMENT MODEL}
\label{sec:linear}

The linear alignment model is the simplest and most widely-used model for intrinsic alignments \citep{catelan}. \altered{This is thought to be applicable to elliptical galaxies, and has had success in modelling the correlations of LRGs which have been measured to high significance, see for example~\citealt{Singh:2014kla}.} It assumes that galaxy shapes follow, to some extent, those of their host dark matter halos, which are tidally sheared by the large-scale structure. 
For a galaxy at distance $\chi$ along the line-of-sight $\hat{\vec{n}}$ this induces an intrinsic ellipticity given by
\begin{equation}
\epsilon^{\rm s}(\hat{\vec{n}};\chi) = -\frac{C}{4\pi G}\frac{1}{\chi^2} \eth^2  S\left[\Phi(\chi\hat{\vec{n}},\chi_{\rm P})\right] \, ,
\label{eqn:catelan1}
\end{equation}
where $S$ represents a smoothing of the linear-theory potential over small scales.\footnote{Typically, the smoothing scale is taken to be around $1\,{\rm Mpc}$, corresponding to size of the dark matter halo. Here we use a cutoff in Fourier space at $k=10\, h\rm{Mpc}^{-1}$. } The potential is evaluated at the time of galaxy formation $\chi_{\rm P}$. The constant $C$ is expected to be positive (so that galaxies are tidally sheared so their long axes are towards the overdensities of a surrounding quadrupole mass distribution) and has the dimensions of inverse density. An improved model can be obtained by replacing the linear Newtonian potential $\Phi$ with its non-linear counterpart~\citep{bridleking}. This is referred to as the \emph{non-linear alignment model}; we briefly summarize the main results of this model for the cross-correlation with CMB lensing in the remainder of this section. More details can be found in~\citet{catelan} and~\citet{hirataseljak04} for galaxy--galaxy cross correlations or \citet{hall} and \citet{iacross2} for CMB--galaxy cross-correlations.

Under the Limber approximation, the cross-correlation between $\epsilon^{\rm s}$ and the CMB lensing potential $\phi$ evaluates to
\begin{equation}
C_l^{\epsilon^{\rm s}\phi} = \frac{2C}{4\pi G} \sqrt{\frac{(l+2)!}{(l-2)!}}\int_0^{\chi_*} d\chi\, \frac{f(\chi )}{\chi^4}\frac{\chi_*-\chi}{\chi \chi_*} \frac{1 }{ \bar{D}(z)}P_{\Phi}(l/\chi; \chi) .
\label{eqn:linearcrosse}
\end{equation}
Here, $\bar{D}(z)\propto (1+z)D(z)$ is the rescaled growth function appropriate for the gravitational potential, and is normalized to unity at high redshifts.\footnote{%
More generally, we approximate the non-linear potential power spectrum between modes at time $\chi$ and $\chi_{\rm P}$ as
\begin{equation*}
P_\Phi(k;\chi,\chi_{\rm P}) \approx \left[P_\Phi(k;\chi) P_\Phi(k;\chi_{\rm P})\right]^{1/2} \, ,
\end{equation*}
and so replace $\bar{D}(z)$ with
\begin{equation*}
\bar{D}(z) \rightarrow \sqrt{\frac{P_\Phi(k;\chi)}{P_\Phi(k;\chi_{\rm P})}} \, ,
\end{equation*}
which is both time- and scale-dependent.}
It accounts for evolution of the potential from the time of galaxy formation $\chi_{\rm P}$ to $\chi$.
For illustrative purposes, we take the constant $C=5\times 10^{-14} h^{-2}\,\rm{M_{\odot}}^{-1}\,\rm{Mpc}^3$ following~\citet{bridleking} who showed that this value matches the SuperCOSMOS survey observations reported in~\citet{brown}. Note that the $C_l^{\epsilon^{\rm s}\phi}$ is positive, and so intrinsic alignments reduce the magnitude of the overall cross-correlation signal in this model.

\begin{figure}
\centering
\includegraphics[width=0.5\textwidth]{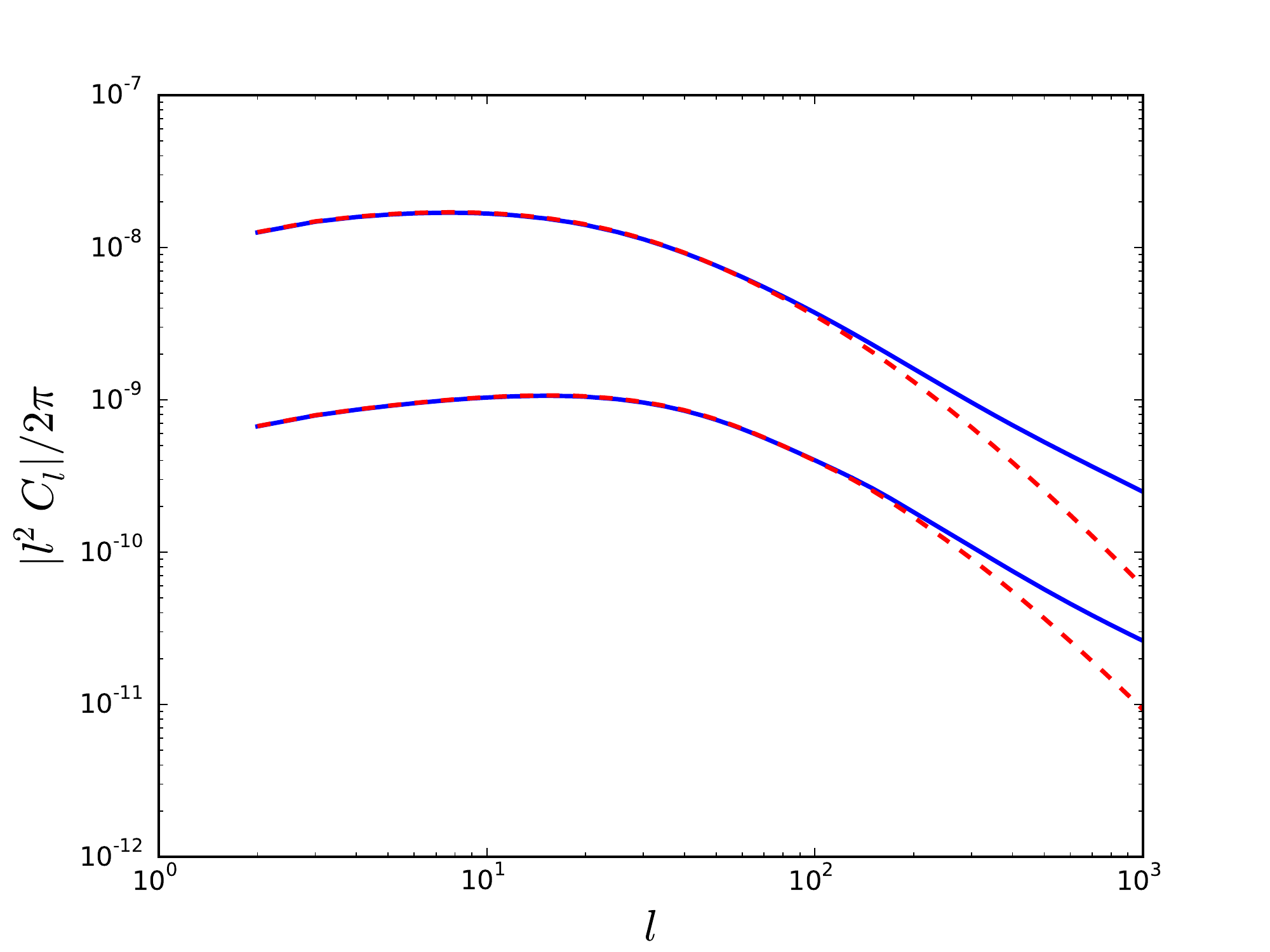}
\noindent
\caption{Absolute value of the cross-spectrum of the CMB lensing potential with the gravitational shear, $C_l^{\gamma\phi}$, (upper lines) calculated with the linear (blue solid) and non-linear (red dashed) matter power spectrum. The correlation between the intrinsic ellipticity and the CMB lensing potential, $C_l^{\epsilon^{\rm s}\phi}$, is also shown (lower lines) in the linear alignment model (red dashed) and the non-linear alignment model (solid blue). The galaxy redshift distribution is given by Eq.~\eqref{eq:CS82}.}
\label{fig:linear}
\end{figure}

Figure~\ref{fig:linear} compares the intrinsic alignment contamination, $C_l^{\epsilon^{\rm s}\phi}$, to the gravitational shear signal, $C_l^{\gamma\phi}$, in the linear and non-linear alignment models for the CS82 redshift distribution from~\citet{hand13}:
\begin{equation}
f(\chi) = A H(z)\frac{z^a + z^{ab}}{z^b +c} \, ,
\label{eq:CS82}
\end{equation}
with $a=0.531$, $b=7.810$, $c=0.517$ and $A$ chosen to normalize the distribution.
In both cases the intrinsic-alignment term reduces the cross-correlation by around 15\,\%, as found by~\citet{hall} and \citet{iacross2}. Non-linear corrections are mild up to multipoles of a few hundred, and it is on these large and intermediate scales where current measurements of CMB lensing have most of their signal to noise.

\section{QUADRATIC ALIGNMENT MODEL}
\label{sec:quad}

The quadratic alignment model is used to describe the alignments of disc galaxies. It is based on tidal-torque theory (see~\citealt{Schaefer:2008xd} for a detailed review), which asserts that a protogalaxy in a tidal field will experience a torque, provided that there is some misalignment between the inertia tensor and the tidal tensor~\citep{1984ApJ...286...38W}. The galactic disc is assumed to form perpendicular to the resulting angular momentum vector, giving projected ellipticities in the flat-sky approximation (with line-of-sight direction along the $z$-direction and the ellipticity components defined with respect to $x$ and $y$) of the form
\begin{align}
\epsilon^{\rm s}_{1} &= f(L,L_z) (L_x^2-L_y^2) \nonumber \\
\epsilon^{\rm s}_{2} &= 2f(L,L_z) L_x L_y \, .
\label{eqn:e2}
\end{align}
Here, $f(L,L_z)$ can depend only on the magnitude of the angular moment, $L$, and the $z$-component by symmetry. Generally, we expect $f(L,L_z)$ to be negative since angular momentum in the $y$--$z$ plane gives the major axis of the projected disc along the $x$-axis. For the case of a thin disc, we have $f(L,L_z) = - 1/(L+|L_z|)^2$ given our definition of ellipticity.
Here, however, we choose to follow~\citet{mwk} by taking $f(L,L_z)$ to be a constant, $-C$. In the final correlation functions, this approximation gives similar results to models that assume galaxies are thin discs (see~\citealt{cnpt}) but greatly simplifies the calculations. 

\altered{The acquisition of angular momentum through tidal torquing is best described in the Lagrangian approach. We denote comoving Eulerian coordinates by $\vec{x}$. For a given fluid element, its Eulerian coordinates at time $t$ are related to its (fixed) Lagrangian coordinates $\vec{q}$ by $\vec{x}(t)=\vec{q}+\vec{S}(\vec{q},t)$, where $\vec{S}(\vec{q},t)$ is the displacement field. The angular momentum at time $t$ of the matter contained within a Lagrangian volume $V_{q}$ can be shown to be~\citep{1984ApJ...286...38W}
\begin{equation}
\vec{L}(t) = \rho_{\rm m}(t)a^5(t) \int_{V_q} \left(\vec{q}-\vec{q}_{\rm COM}\right)\times
\dot{\vec{x}}(t) \, d^3 \vec{q}  
\end{equation}
correct to second order in perturbations, where $\vec{q}_{\rm COM}$ is the Lagrangian centre of mass. At first and second order, the displacement $\vec{S}(\vec{q},t)$ can be written as the gradient of a displacement potential. Approximating the cosmology as Einstein--de Sitter over the epoch of galaxy formation, we can write
\begin{equation}
\vec{S}(\vec{q},t) = a \vec{\nabla}_{\vec{q}} \psi^{(1)}(\vec{q}) + a^2 \vec{\nabla}_{\vec{q}}\psi^{(2)}(\vec{q}) + \cdots \, ,
\end{equation}
where $\psi^{(1)}(\vec{q})$ and $\psi^{(2)}(\vec{q})$ are the first- and second-order displacement potentials, respectively. In terms of these, the Eulerian peculiar velocity to second order is
\begin{equation}
\dot{x}(\vec{q},t) = a H \vec{\nabla}_{\vec{q}} \psi \, ,
\end{equation}
where $H=\dot{a}/a$ is the Hubble parameter and
\begin{equation}
\psi(\vec{q},t) \equiv \psi^{(1)}(\vec{q}) +  2 a \psi^{(2)}(\vec{q})
\end{equation}
is the velocity potential. Finally, Taylor expanding $\psi$ about $\vec{q}_{\rm COM}$, we find
\begin{equation}
L_i = a^3 H \epsilon_{ijk} I_{jl} \left. \frac{\partial^2 \psi}{\partial q_k \partial q_l} \right|_{\vec{q}_{\rm COM}}
\label{eq:Lgen}
\end{equation}
correct to second order. Here,
\begin{equation}
I_{ij} \equiv \rho_{\rm m} a^3 \int_{V_q} \left(q-q_{\rm COM}\right)_i
\left(q-q_{\rm COM}\right)_j \, d^3 \vec{q}
\end{equation}
is the (constant) Lagrangian inertia tensor and $\epsilon_{ijk}$ is the alternating tensor.
}

\altered{We see from Eq.~(\ref{eq:Lgen}) that misalignment between the Lagrangian inertia tensor and the tidal tensor formed from the velocity potential leads to the acquisition of angular momentum. Assuming an axisymmetric inertia tensor, with symmetry axis $\hat{\vec{m}}$, the angular momentum of the protogalaxy is given by
\begin{equation}
L_i \propto \epsilon_{ijk} \frac{\partial^2 \psi}{\partial q_k \partial q_l}
\hat{m}_l \hat{m}_k \, ,
\label{eqn:Li}
\end{equation}
where $\psi(\vec{q},t)$ is evaluated at the time of galaxy formation and at $\vec{q}_{\rm COM}$. Following~\citet{mwk}, we assume that the inertia tensor of the protogalaxy is independent of the local tidal field (see e.g.,~\citealt{cnpt} for a relaxation of this assumption). In this case, the expected intrinsic ellipticity for a given tidal field is obtained by averaging over the direction $\hat{\vec{m}}$, so that Eqs.~\eqref{eqn:e2} and~\eqref{eqn:Li} give
\begin{align}
\epsilon^{\rm s}_1 &= \frac{3C}{15} \sum_{i=1}^3 K_{1i}K_{1i} - K_{2i} K_{2i} \nonumber \\
\epsilon^{\rm s}_2 &= \frac{6C}{15} \sum_{i=1}^3 K_{1i} K_{2i} \, .
\label{eq:qeps}
\end{align}
Here, the trace-free tidal tensor
\begin{equation}
K_{ij} \equiv D_{ij} \psi = \frac{\partial^2 \psi}{\partial q_i \partial q_j}
- \frac{1}{3} \delta_{ij} \nabla_{\vec{q}}^2 \psi \, .
\end{equation}}
The quantity $C$ is a positive normalisation constant; below we shall relate it to the dispersion of the intrinsic ellipticity.


We emphasise that the quadratic alignment model is likely a severe simplification of the physics involved. Small-scale phenomena such as vorticity around filaments, mergers and gas accretion will alter the alignments, and the angular momentum of the galaxy and its host halo will have some misalignment. However, some of the key ideas on which the model is built have been observed in data and reproduced in simulations. For example, the quadratic dependence of the alignments on the gravitational potential has been verified observationally for bright blue galaxies at low redshift~\citep{Lee:2007tq}, and \citet{Codis:2014awa} recently found that blue galaxies in the Horizon-AGN simulation remain correlated with the tidal field at a redshift of $z=1.2$. 

%
%
In this section, we calculate the correlation between $\epsilon^{\rm s}$ and the CMB lensing potential in the flat-sky approximation, extending the calculation of the auto-correlations of $\epsilon^{\rm s}$ from~\citet{mwk}. A related calculation is given for the correlation between $\epsilon^{\rm s}$ and the galaxy overdensity by~\citet{huizhang}.

The flat-sky analogue of the spin-2 expansion of $\epsilon^{\rm s}$ into $E$ and $B$-modes,
\begin{equation}
\epsilon_\pm^{\rm s}(\hat{\vec{n}}) = \sum_{lm} \left(\epsilon^{\rm s}_{E, lm} \pm i \epsilon^{\rm s}_{B, lm}\right) {}_{\pm 2}Y_{lm}(\hat{\vec{n}}) \, ,
\end{equation}
where $\epsilon^{\rm s}_\pm = \epsilon^{\rm s}_1 \pm i \epsilon^{\rm s}_2$, is
\begin{equation}
\epsilon_\pm^{\rm s}(\vec{\theta}) = - \int \frac{d^2 \vec{l}}{2\pi} \, \left[\epsilon^{\rm s}_E(\vec{l}) \pm i \epsilon^{\rm s}_B(\vec{l}) \right] e^{\pm 2 i \phi_{\vec{l}}} e^{i \vec{l}\cdot\vec{\theta}} \, .
\end{equation}
Here, $\vec{\theta}$ is the angular position in the plane of the sky, $l_x \pm i l_y = l e^{\pm i \phi_{\vec{l}}}$ with $l = |\vec{l}|$, and $\epsilon_\pm^{\rm s}(\vec{\theta})$ are defined with respect to the $x$- and $y$-directions. Since the $B$-modes do not correlate with the scalar lensing potential $\phi$, the cross power spectrum follows from
\begin{equation}
\langle \epsilon^{\rm s}_E(\vec{l}) \phi^\ast(\vec{l}') \rangle = \delta^{(2)}(\vec{l}-\vec{l}') C_l^{\epsilon^{\rm s}\phi} \, ,
\end{equation}
where $\phi(\vec{l})$ is the flat-sky Fourier transform of the lensing potential.

For a source at distance $\chi$, $\epsilon^{\rm s}_E(\vec{l}) \pm i \epsilon^{\rm s}_B(\vec{l})$ depends on the 3D Fourier transform $\epsilon_\pm^{\rm s}(\vec{k},\chi)$ at lookback time $\chi$ and wavevector perpendicular to the line of sight $\vec{k}_\perp = \vec{l}/\chi$. \altered{There is a subtlety in the calculation of $\epsilon_\pm^{\rm s}(\vec{k})$ since the Fourier transform is with respect to the Eulerian position of the galaxy at lookback time $\chi$, i.e., it involves the components of
\begin{equation}
\mathcal{K}_{ij}(\vec{k}) = \int \frac{d^3 \vec{x}}{(2\pi)^{3/2}} e^{-i \vec{k} \cdot
\vec{x}(\vec{q},t)} K_{ik}(\vec{q},t_{\rm P}) K_{kj}(\vec{q},t_{\rm P}) \, ,
\end{equation}
where $t$ corresponds to lookback time $\chi$. Here, we have approximated the Eulerian centre-of-mass position of the galaxy at time $t$ as $\vec{x}_{\rm COM}(t)  = \vec{x}(\vec{q}_{\rm COM},t)$. We require $\mathcal{K}_{ij}(\vec{k})$ to third order to calculate the leading-order non-zero correlation with the lensing potential $\phi$. Changing variables from $\vec{x}$ to $\vec{q}$, and expanding, we have
\begin{multline}
\mathcal{K}_{ij}(\vec{k}) = \int \frac{d^3 \vec{q}}{(2\pi)^{3/2}} e^{-i \vec{k} \cdot \vec{q}} K_{ik}(\vec{q},t_{\rm P}) K_{kj}(\vec{q},t_{\rm P}) \\
\times \left(1 + a \nabla^2_{\vec{q}}\psi^{(1)} - i a \vec{k}\cdot \nabla_{\vec{q}} \psi^{(1)} \right) \, ,
\label{eq:calKij}
\end{multline}
correct to third order. Here, the term $a \nabla^2_{\vec{q}}\psi^{(1)}$ comes from the Jacobian of the transformation between $\vec{x}$ and $\vec{q}$. The terms involving $\psi^{(1)}$ complicate the calculation of $\mathcal{K}_{ij}(\vec{k})$; however, they do not enter the correlation of $\epsilon_\pm^{\rm s}(\vec{k},\chi)$ with the lensing potential. To see this, consider the correlation of these terms with the gravitational potential $\Phi$. We only need use linear theory to evaluate these correlations to fourth order in perturbations. On taking the expectation value, there are three types of Gaussian contractions to consider. Those involving $\langle K_{ik}(\vec{q},t_{\rm P}) K_{kj}(\vec{q},t_{\rm P}) \rangle \propto \delta_{ij}$ do not contribute to $\epsilon_\pm^{\rm s}(\vec{k},\chi)$ by statistical isotropy. The second type involves $\langle \nabla^2_{\vec{q}} \psi^{(1)} (\vec{q},t) K_{ij}(\vec{q},t_{\rm P})\rangle = 0$, since $K_{ij}$ is trace-free. Finally, we have terms involving the correlation of $K_{ij}$ with $\partial \psi^{(1)} / \partial q_k$ at the same point, but these also vanish by statistical isotropy.
}

The $E$-modes $\epsilon^{\rm s}_E(\vec{l})$ depend only on
\begin{align}
\epsilon^{\rm s}_E(\vec{k}) &\equiv - \frac{1}{2} \left[ \epsilon^{\rm s}_+(\vec{k}) e^{-2i \phi_{\vec{k}_\perp}}+ \epsilon^{\rm s}_-(\vec{k}) e^{2i \phi_{\vec{k}_\perp}} \right] \nonumber \\
&= - \frac{1}{2 \vec{k}_\perp^2} \left[ (k_x^2-k_y^2) \epsilon^{\rm s}_1(\vec{k}) + 2 k_x k_y \epsilon^{\rm s}_2(\vec{k}) \right] \, ,
\end{align}
which transform as a scalar under rotations about the $z$-axis. Ultimately, in the Limber approximation, only Fourier modes with $\vec{k}$ perpendicular to the $z$-direction contribute to the correlation functions. Using Eq.~\eqref{eq:qeps}, and retaining only those terms from Eq.~\eqref{eq:calKij} that contribute to the correlation with the lensing potential, we find~\citep{mwk}
\begin{multline}
\vec{k}_\perp^2 \epsilon^{\rm s}_E(\vec{k}) \supset \frac{C}{15} \int \frac{d^3 \vec{k}' d^3 \vec{k}''}{(2\pi)^{3/2}} \, \bigl[ f_E(\vec{k}',\vec{k}'') \\
\times \psi(\vec{k}',\chi_{\rm P})
\psi(\vec{k}'',\chi_{\rm P}) \delta^{(3)}(\vec{k}'+\vec{k}''-\vec{k}) \bigr] \, ,
\end{multline}
where
\begin{align}
2f_{E}(\vec{k}',\vec{k}'') &=  \left(2k_z^{\prime 2}-\vec{k}_\perp^{\prime 2}\right)\left[\left(\vec{k}'_\perp \cdot \vec{k}''_\perp + \vec{k}_\perp^{\prime\prime 2}\right)^2 - \left(\vec{k}'_\perp \times \vec{k}''_\perp\right)^2 \right] \nonumber \\
&\mbox{}+ \left(2k_z^{\prime\prime 2}-\vec{k}_\perp^{\prime\prime 2}\right)\left[\left(\vec{k}'_\perp \cdot \vec{k}''_\perp + \vec{k}_\perp^{\prime 2}\right)^2 - \left(\vec{k}'_\perp \times \vec{k}''_\perp\right)^2 \right] \nonumber \\
&\mbox{} -6 k'_z k''_z\left[ \vec{k}'_\perp \cdot \vec{k}''_\perp \left(\vec{k}_\perp^{\prime 2}+\vec{k}_\perp^{\prime\prime 2}\right) + 2 \vec{k}_\perp^{\prime 2} \vec{k}_\perp^{\prime\prime 2} \right] \, .
\end{align}
Note that $f_E$ is invariant under rotations of its arguments about the $z$-direction.
The (dimensional) 3D cross power spectrum between $\epsilon^{\rm s}_E(\vec{k})$ and $\Phi(\vec{k})$ is defined by
\begin{equation}
\langle \epsilon^{\rm s}_E(\vec{k}) \Phi(\vec{k}') \rangle = \delta^{(3)}(\vec{k}+\vec{k}')
P_{\epsilon^{\rm s}_E \Phi}(|\vec{k}_\perp|,k_z) \, ,
\end{equation}
and depends only on the magnitude of $\vec{k}_\perp$ and $k_z$. Since $\epsilon^{\rm s}_E(\vec{k})$ is quadratic in the velocity potential $\psi$, the cross power spectrum involves the $\psi\psi\Phi$ bispectrum, defined by (with our Fourier conventions)
\begin{multline}
\langle \psi(\vec{k}_1,\chi_1) \psi(\vec{k}_2, \chi_2) \Phi(\vec{k}_3, \chi_3) \rangle
= (2\pi)^{-3/2} \delta^{(3)}(\vec{k}_1+\vec{k}_2+\vec{k}_3) \\
\times B_{\psi\psi\Phi}(k_1,k_2,k_3;\chi_1,\chi_2,\chi_3) \, .
\end{multline}
It follows that
\begin{multline}
\vec{k}_\perp^2 P_{\epsilon^{\rm s}_E \Phi}(|\vec{k}_\perp|,k_z) = \frac{C}{15} 
\int \frac{d^3 \vec{k}' d^3 \vec{k}''}{(2\pi)^{3/2}} \, \bigl[ f_E(\vec{k}',\vec{k}'') \\
\times B_{\psi\psi\Phi}(k',k''\!,k;\chi_{\rm P},\chi_{\rm P},\chi) \delta^{(3)}(\vec{k}'+\vec{k}''-\vec{k}) \bigr] \, ,
\label{eq:Pepsphi}
\end{multline}
Finally, in the Limber approximation and for a source distribution $f(\chi)$, the cross-correlation between the intrinsic alignments and the CMB lensing potential evaluates to
\begin{equation}
C_l^{\epsilon^{\rm s}\phi} = -2 \int_0^{\chi_\ast} d\chi\, \frac{f(\chi)}{\chi^2} \frac{\chi_\ast - \chi}{\chi_\ast\chi} P_{\epsilon^{\rm s}_E \Phi}(l/\chi,k_z=0) \, ,
\end{equation}
where, for simplicity, we have assumed that all galaxies form at the same time $\chi_{\rm P}$.

We evaluate $P_{\epsilon^{\rm s}_E \Phi}(|\vec{k}_\perp|,k_z=0)$ from Eq.~\eqref{eq:Pepsphi}
by taking $\vec{k} = k(1,0,0)$ and writing $\vec{k}'$ in terms of a magnitude $k' = \alpha k$, a polar angle $\theta$ and an azimuthal angle $\varphi$:
\begin{equation}
\vec{k}' = \alpha k (\cos\theta,\sin\theta \cos\varphi,\sin\theta \sin\varphi) \, .
\end{equation}
In Eq.~\eqref{eq:Pepsphi}, the delta-function enforces $\vec{k}'' = \vec{k} - \vec{k}'$, which, with the above parameterisation, has magnitude
\begin{equation}
k'' = k \sqrt{1 + \alpha^2 - 2 \alpha \mu} \, ,
\label{eq:kpp}
\end{equation}
where $\mu \equiv \cos\theta$,
so the bispectrum term in the integrand of Eq.~\eqref{eq:Pepsphi} depends only on $k$, $\alpha$ and $\mu$. We can therefore integrate $f_E(\vec{k}',\vec{k}'')$ over $\varphi$ to find
\begin{align}
\int \frac{d\varphi}{2\pi} \, f_E(\vec{k}',\vec{k}'') &= \alpha^2 k^6 \frac{1}{2} \left[
(1+\alpha^2)(1-3\mu^2)+\alpha\mu(1+3\mu^2) \right] \nonumber \\
&= \alpha^2 k^6 \bar{f}_E(\alpha,\mu) \, ,
\end{align}
so that
\begin{multline}
P_{\epsilon^{\rm s}_E \Phi}(|\vec{k}_\perp|,k_z=0) = \frac{C}{15} \int \frac{k^{\prime 2} dk' d\mu}{(2\pi)^2} \bigl[ \alpha^2 k^4 \bar{f}_E(\alpha,\mu) \\
\times B_{\psi\psi\Phi}(k',k''\!,k;\chi_{\rm P},\chi_{\rm P},\chi) \bigr] \, ,
\end{multline}
where $\alpha = k'/k$ and $k''$ is given by Eq.~\eqref{eq:kpp}. Using this in
Eq.~\eqref{eq:Pepsphi}, we obtain our final expression for the cross power spectrum:
\begin{multline}
C_l^{\epsilon^{\rm s}\phi} = -\frac{2C}{15} \int_0^{\chi_\ast} d\chi\, \frac{f(\chi)}{\chi^2} \frac{\chi_\ast - \chi}{\chi_\ast\chi} \int 
\frac{k^{\prime 2} dk' d\mu}{(2\pi)^2} \bigl[ \alpha^2 k^4 \\
\times \bar{f}_E(\alpha,\mu)
B_{\psi\psi\Phi}(k',k''\!,k;\chi_{\rm P},\chi_{\rm P},\chi) \bigr] \, ,
\label{eqn:crosscorr}
\end{multline}
with $k = l / \chi$.

We now consider the normalisation constant $C$. Following \citet{mwk} again, we determine this by assuming that all of the intrinsic ellipticity of galaxies comes from tidal torquing (so we likely overestimate the amplitude of the intrinsic-alignment correlations). In this case, we equate the mean-squared intrinsic ellipticity of a single galaxy to the squared dispersion of the intrinsic ellipticity over the galaxy population, $\epsnorm$, to find
\begin{equation}
\epsnorm= \langle (\epsilon_1^{\rm s})^2 + (\epsilon_2^{\rm s})^2 \rangle = C^2 \langle (L_x^2 + L_y^2)^2 \rangle \, .
\end{equation}
The average here is over the direction $\hat{\vec{m}}$ of the symmetry axis of the inertia tensor and over realisations of $\psi$ (which are assumed independent). At fixed $\hat{\vec{m}}$, the angular momentum is proportional to $\psi$ and so
\begin{equation}
\langle L_i L_j \rangle_\psi = \frac{C}{15} \langle (\nabla_{\vec{q}}^2 \psi)^2\rangle (\delta_{ij} - \hat{m}_i \hat{m}_j) \, ,
\end{equation}
where the expectation value is over $\psi$.
Approximating $\psi$ as a Gaussian random field, we use this result to take the expectation value of $(L_x^2 + L_y^2)^2$ at fixed $\hat{\vec{m}}$ using Wick's theorem, and finally average over $\hat{\vec{m}}$. The end result is\footnote{Equation~\eqref{eq:Cnorm} corrects the result in~\citet{mwk}; their Eq.~(24) differs by a factor of $\sqrt{3/2}$.}
\begin{equation}
C = \frac{15^{3/2}}{8}\frac{\sqrt{\epsnorm}}{\langle (\nabla_{\vec{q}}^2 \psi)^2\rangle} \, ,
\label{eq:Cnorm}
\end{equation}
where $\psi$ is evaluated at the time of galaxy formation.


\subsection{Bispectrum}

\altered{In the quadratic alignment model, the cross-correlation between the intrinsic ellipticity and the CMB lensing potential depends on the bispectrum of the velocity and gravitational potentials, and so vanishes for Gaussian potentials.\footnote{Note that this holds true even if source clustering corrections are included for the galaxy lensing~\citep{hirataseljak04,huizhang}.} In particular, Eq.~\eqref{eqn:crosscorr} involves the two-time bispectrum $B_{\psi\psi\Phi}(k_1,k_2,k_3;\chi_{\rm P},\chi_{\rm P},\chi)$. We evaluate this at tree-level using standard results from second-order perturbation theory (e.g.,~\citealt{1995A&A...296..575B}). For the displacement potential, we have
\begin{equation}
\psi^{(2)}(\vec{k}) = \frac{1}{2}\int \frac{d^3\vec{k}_1}{(2\pi)^{3/2}} \,
\frac{k_1^2 k_2^2}{k^2} G(\vec{k}_1,\vec{k}_2) \psi^{(1)}(\vec{k}_1) \psi^{(1)}(\vec{k}_2) \, ,
\label{eq:psi2kspace}
\end{equation}
where $\vec{k}_2 = \vec{k}-\vec{k}_1$ and the Lagrangian coupling kernel
\begin{equation}
G(\vec{k}_1,\vec{k}_2) =  \frac{3}{7}-\frac{3}{7}\left(\frac{\vec{k}_1\cdot \vec{k}_2}{k_1 k_2}\right)^2 \, .
\end{equation}
For the gravitational potential, 
\begin{equation}
\Phi(\vec{k},t) = \Phi^{(1)}(\vec{k},t) + \Phi^{(2)}(\vec{k},t) + \cdots \, , 
\end{equation}
at linear order
\begin{equation}
\Phi^{(1)}(\vec{k},t) = -\frac{3}{2} \left(a^3 H^2\right) \psi^{(1)}(\vec{k})
\end{equation}
in an Einstein--de Sitter universe. More generally, one replaces $3 a^3 H^2 /2$ with
$4\pi G a^3 \rho_{\rm m}\bar{D}(a)$ where, recall, $\bar{D}(a)$ is the rescaled linear growth function normalised to unity as $a\rightarrow 0$. At second order,
\begin{multline}
\Phi^{(2)}(\vec{k},t) \approx -4\pi G a^4 \rho_{\rm m} \bar{D}^2 
\int \frac{d^3\vec{k}_1}{(2\pi)^{3/2}} \,
\frac{k_1^2 k_2^2}{k^2} F(\vec{k}_1,\vec{k}_2) \\
\times \psi^{(1)}(\vec{k}_1) \psi^{(1)}(\vec{k}_2) \, ,
\end{multline}
where we have made the (excellent) approximation that the growth rate for the second-order displacement potential is the square of that for the first-order potential.
Here, the Eulerian coupling kernel $F(\vec{k}_1,\vec{k}_2)$ is given by
\begin{equation}
F(\vec{k}_1,\vec{k}_2) = \frac{5}{7} + \frac{\vec{k}_1\cdot\vec{k}_2}{2 k_1 k_2}\left(\frac{k_1}{k_2}+\frac{k_2}{k_1}\right) + \frac{2}{7} \left(\frac{\vec{k}_1\cdot \vec{k}_2}{k_1 k_2}\right)^2 \, .
\end{equation}
Using these results, we find
\begin{multline}
B_{\psi\psi\Phi}(k_1,k_2,k_3; \chi_P, \chi_P, \chi) = - 8\pi G a^3 \rho_{\rm m} \\
\times 
\left(\frac{k_1^2 k_2^2}{k_3^2} a(\chi) \bar{D}^2(\chi) P_{\psi^{(1)}}(k_1) P_{\psi^{(1)}}(k_2) 
F(\vec{k}_1,\vec{k}_2) \right. \\
\hspace{0.35cm} + \frac{k_1^2 k_3^2}{k_2^2} a(\chi_P) \bar{D}(\chi) P_{\psi^{(1)}}(k_1) P_{\psi^{(1)}}(k_3) 
G(\vec{k}_1,\vec{k}_3)\\
 + \left. \frac{k_2^2 k_3^2}{k_1^2} a(\chi_P) \bar{D}(\chi) P_{\psi^{(1)}}(k_2) P_{\psi^{(1)}}(k_3) 
G(\vec{k}_2,\vec{k}_3) \right) \, ,
\label{eq:treebispectrum}
\end{multline}
approximating the cosmology as Einstein--de Sitter at $\chi_P$. Here, $P_{\psi^{(1)}}(k)$ is the (time-independent) power spectrum of the linear-theory displacement potential $\psi^{(1)}(\vec{k})$, which is proportional to the power spectrum of the linear-theory gravitational potential $\Phi(\vec{k})$ at early times.}

The various terms in Eq.~\eqref{eq:treebispectrum} have different dependencies on $\chi_{\rm P}$ and $\chi$. As we discuss in Sec.~\ref{subsec:squeezed}, the second and third terms dominate the cross-correlation on large and intermediate scales. In the bispectrum, these terms scale as $a(\chi_{\rm P})$ implying a strong dependence of the cross-correlation on the (uncertain) time of galaxy formation, with earlier times giving a smaller correlation.

\altered{In practice, we use an extension of the tree-level bispectrum~\eqref{eq:treebispectrum}, suggested by~\citet{fastestimation} in the context of the matter bispectrum where it better describes the weakly non-linear regime. In this approximation, the structure of the tree-level result is retained but the effect of loops on the external legs are accounted for by using the non-linear $\Phi$ power spectra in place of their linear-theory $\psi$ counterparts.\footnote{ \altered{In detail, in the first term in~\eqref{eq:treebispectrum}, we replace $\bar{D}^2(\chi) P_{\psi^{(1)}}(k_1) P_{\psi^{(1)}}(k_2)$ with the non-linear $P_\Phi(k_1;\chi_P,\chi) P_\Phi(k_2;\chi_P,\chi)$ (divided by the square of $4\pi G a^3 \rho_{\rm m}$); in the second term we replace
$\bar{D}(\chi) P_{\psi^{(1)}}(k_1) P_{\psi^{(1)}}(k_3)$ with the non-linear $P_\Phi(k_1;\chi_P) P_\Phi(k_3;\chi_P,\chi)$ and similarly in the third term (but with $k_1$ replaced by $k_2$).} Note that this procedure has not been tested against numerical simulations for the $B_{\psi\psi\Phi}$ bispectrum that we require here.}
The non-linear potential power spectra are calculated from the non-linear matter power spectrum via the Poisson equation in Fourier space.
We use the HALOFIT~\citep{Smith:2002dz} fitting function implemented in the CAMB code~\citep{Lewis:1999bs}. 
}

\subsection{Squeezed-limit approximation}
\label{subsec:squeezed}

Physically, we expect the cross-correlation~\eqref{eqn:crosscorr} on large and intermediate scales to be dominated by the squeezed limit of the bispectrum with $k' \approx k'' \gg k$. Since the intrinsic ellipticity depends quadratically on the tidal tensor, it is dominated by small-scale modes with wavenumbers up to the assumed smoothing scale. The squeezed limit of the bispectrum corresponds to a modulation of the power in small-scale modes by the large-scale modes responsible for the CMB lensing. This modulates the intrinsic ellipticities correlating them with the CMB mode. (See~\citealt{baldauf} and \citealt{BAOshift} for similar arguments in other applications.) We verify the dominance of the squeezed limit numerically in Sec.~\ref{subsec:results}.

\altered{In the squeezed limit, we can simplify the tree-level bispectrum further. Writing it as a function of $k$, $\alpha$ and $\mu$, and expanding to leading order in $1/\alpha$, we have
\begin{multline}
B_{\psi\psi\Phi}(k',k'',k;\chi_P,\chi_P,\chi) \approx - 8\pi G a^3 \rho_{\rm m} \\
\times
\frac{6}{7} a(\chi_P)\bar{D}(\chi) k^2 (1-\mu^2) P_{\psi^{(1)}}(k)P_{\psi^{(1)}}(k') \, .
\label{eq:squeezedtreeb}
\end{multline}
This receives approximately equal contributions from the second and third terms on the right of~\eqref{eq:treebispectrum}, while the first term is suppressed by a factor $\alpha^2 P_{\psi^{(1)}}(k')/P_{\psi^{(1)}}(k) \ll 1$.
Finally, we use Eq.~\eqref{eq:squeezedtreeb} in Eq.~\eqref{eqn:crosscorr} and integrate over $\mu$. Retaining the leading term in $1/\alpha$ in the resulting integrand, we find the approximate result
\begin{multline}
C_l^{\epsilon^{\rm s}\phi} \approx 2  l^2 C \left. \langle (\nabla^2_{\vec{q}} \psi)^2 \rangle \right|_{\chi_{\rm P}} \frac{8}{525} 
\left.\left(\frac{2}{3 a^2 H^2}\right)\right|_{\chi_{\rm P}} \\
\times \int_0^{\chi_*} d\chi\, \frac{f(\chi )}{\chi^4}\frac{\chi_*-\chi}{\chi \chi_*} \frac{1 }{ \bar{D}(z)}P_{\Phi}(l/\chi; \chi) .
\label{eqn:quadapprox}
\end{multline}}
Note that this has exactly the same $l$-dependence as the cross-correlation in the linear alignment model.\footnote{\altered{The original version of this paper approximated the tidal term in Eq.~\eqref{eq:Lgen} with the Eulerian tidal tensor formed from the non-linear Eulerian gravitational potential $\Phi$. We have now self-consistently included all second-order terms within tidal-torque theory, leading to a change in the sign and a reduction in the magnitude of the cross-correlation between $\epsilon^{\rm s}$ and $\phi$.}}
It may not be obvious from the calculation above why the shape of the cross-correlation in the squeezed limit of the quadratic alignment model should be the same as in the linear alignment model. The reason is that, in the squeezed limit, we care about the effect of a large-scale mode on the evolution of small-scale modes. 
\altered{We show in Appendix~\ref{app:condexp} that if one averages the intrinsic ellipticity given in Eq.~\eqref{eq:qeps} over small-scale modes of $\psi^{(1)}(\vec{q})$, in the presence of fixed large-scale modes of $\psi^{(1)}(\vec{q})$, in second-order perturbation theory one finds
\begin{equation}
\langle\epsilon^{\rm s}(\hat{\vec{n}};\chi)\rangle_{\rm S} = - \frac{8 C}{525\chi^2} \left.\left( \frac{2}{3 a^2 H^2}\right)\right|_{\chi_{\rm P}}  \langle (\nabla_{\vec{q}}^2 \psi^{(1)}_{\rm S})^2\rangle 
\eth^2 \Phi^{(1)}_{\rm L}\, ,
\label{eq:realspace}
\end{equation}
where $\psi^{(1)}_{\rm S}$ are the short modes of $\psi^{(1)}(\vec{q})$ and $\Phi^{(1)}_{\rm L}$ are the large-scale modes of the Eulerian gravitational potential at lookback time $\chi_{\rm P}$ and evaluated at distance $\chi$ back along the line of sight.}
In the quadratic alignment model, the effect of the large-scale modes on a galaxy's intrinsic shape has, after averaging over small-scale modes, exactly the same structure as in the linear alignment model, i.e., it depends linearly on the large-scale tidal tensor. Correlating Eq.~\eqref{eq:realspace} with the CMB lensing potential, using the Limber approximation, we recover the correlation in Eq.~\eqref{eqn:quadapprox}.

\subsection{Results}
\label{subsec:results}
\begin{figure}
\centering
\includegraphics[width=0.5\textwidth]{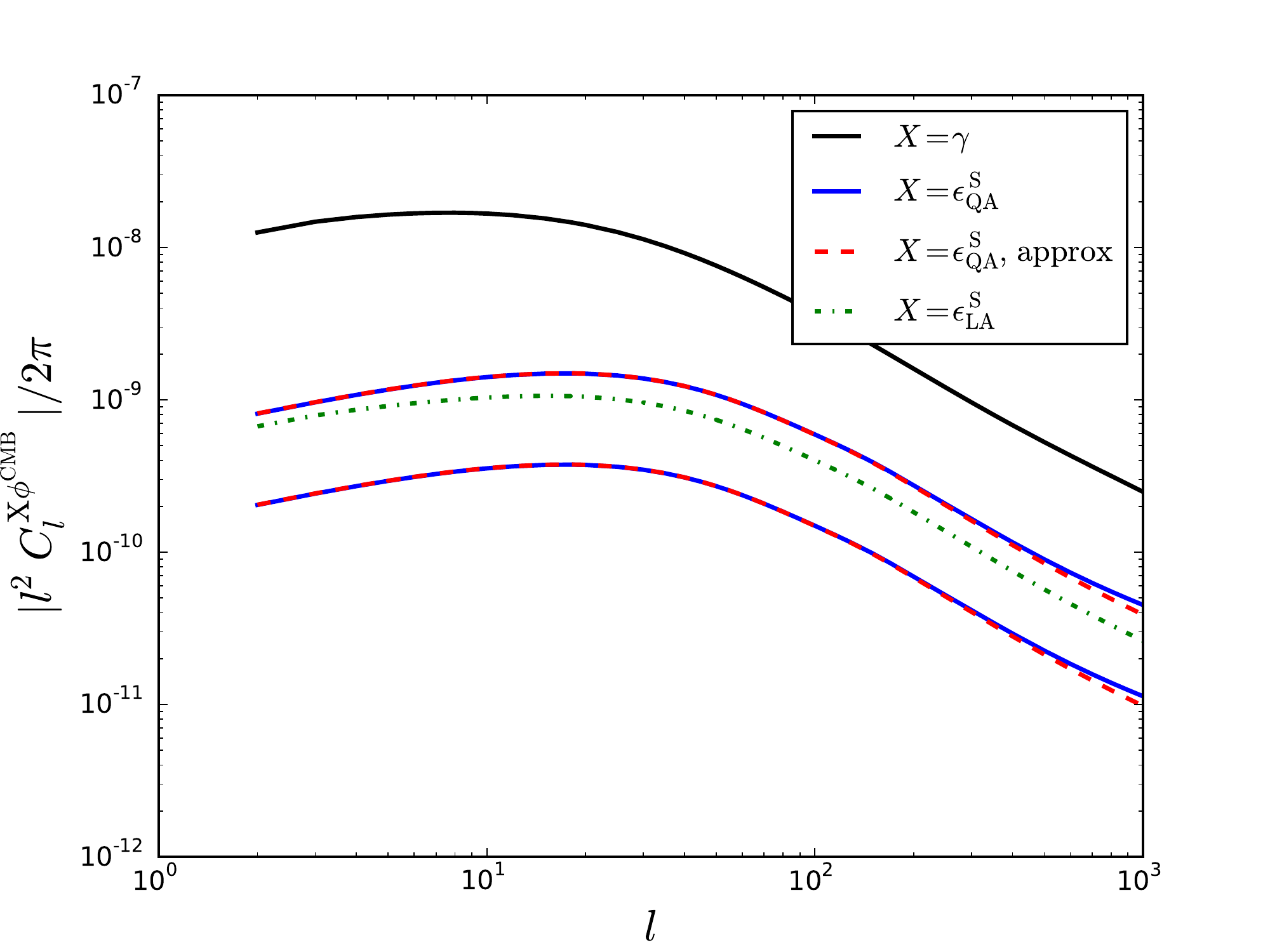}
\noindent
\caption{Absolute value of $C_l^{\epsilon^{\rm s}\phi}$ for the quadratic alignment model (blue solid lines) and the squeezed-limit approximation (red dashed lines) assuming a formation redshift of $z_{\rm P}=3$ (upper lines) and $z_{\rm P}=15$ (lower lines). We also show $C_l^{\gamma\phi}$ (black solid line) and the linear alignment model $C_l^{\epsilon^{\rm s}\phi}$ (green dotted line) for comparison. All spectra are calculated with a cutoff of $k=10\, h{\rm Mpc}^{-1}$ and the redshift distribution given by Eq.~\eqref{eq:CS82}.
}
\label{fig:qi}
\end{figure}

We calculate $C_l^{\epsilon^{\rm s}\phi}$ using the full result from Eqs.~\eqref{eqn:crosscorr} and~\eqref{eq:treebispectrum} (with the non-linear power spectrum), and with the squeezed limit approximation given by Eq.~\eqref{eqn:quadapprox} \altered{for a population of tidally-torqued galaxies with the CS82 redshift distribution} of Eq.~(\ref{eq:CS82}). We take the mean-squared intrinsic ellipticity $\epsnorm = 0.03$ \altered{from \citet{rmse}}. The results are shown, along with the gravitational shear contribution $C_l^{\gamma\phi}$ and $C_l^{\epsilon^{\rm s}\phi}$ for the linear alignment model, in Fig.~\ref{fig:qi}.

\altered{As well as being dependent on the assumed value of the mean-squared intrinsic ellipticity,} the result for the quadratic alignment model contamination is also strongly dependent on the time of galaxy formation, going as $(1+z_{\rm P})^{-1}$ in the squeezed-limit approximation. At high $z_{\rm P}$, the bispectrum of the gravitational potential from non-linear growth is suppressed and the resulting correlation $C_l^{\epsilon^{\rm s}\phi}$ is small. We show results for the endpoints of the plausible range $z_{\rm P} = 3$--15.
The lack of a model of the galaxy formation time allows a wide range of possible values for the contamination of the cross-correlation by intrinsic alignments:
2--7\,\% contamination for the smallest multipoles and 5--18\,\%
 for the highest for the range $z_{\rm P} = 3$--15.
As expected, the shapes of $C_l^{\epsilon^{\rm s}\phi}$ are very similar for the quadratic and the linear alignment models. The amplitudes are also comparable, although we emphasise that our results for the quadratic alignment model should be seen only as upper limits given the assumptions that all alignments and the total intrinsic ellipticity are determined by tidal torquing at the time of galaxy formation.
Finally, we see that the squeezed-limit approximation from Eq.~\eqref{eqn:quadapprox} matches well with the
full result on large and intermediate scales.

\subsection{Comparison with the GI term for galaxy lensing}
\begin{figure}
\centering
\includegraphics[width=0.5\textwidth]{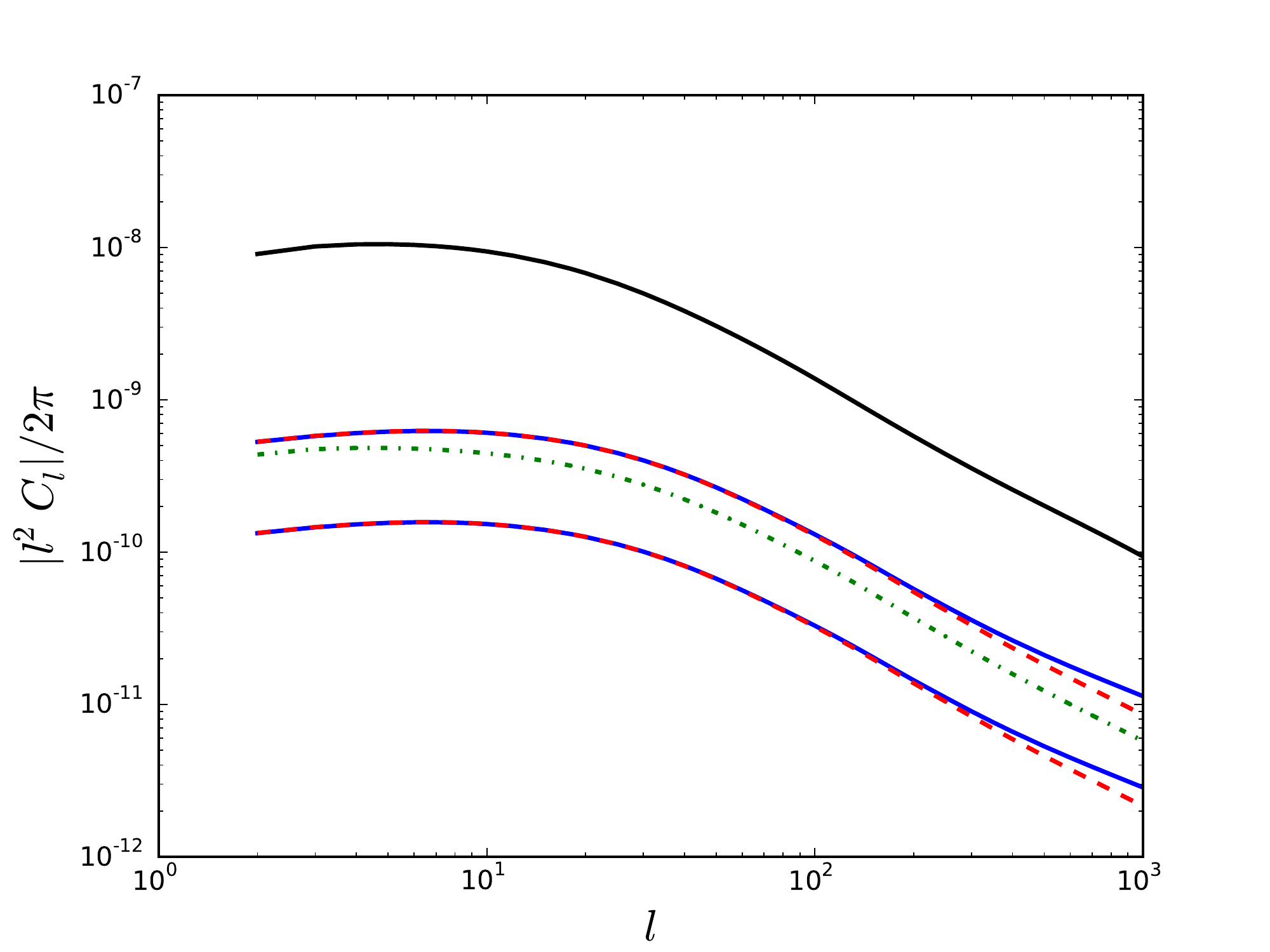}
\noindent
\caption{As Fig.~\ref{fig:qi}, but with the CMB lensing potential replaced by the galaxy lensing potential, $\phi_{\rm gal}$. Note that the total GI contribution to the observable correlation is twice that plotted.}
\label{fig:gi}
\end{figure}

The calculation above can easily be modified to calculate the GI correction to the shear auto-correlation for lensing of galaxies. The gravitational shear $\gamma(\hat{\vec{n}})$ for galaxies with redshift distribution $f(\chi)$ is derivable from a lensing potential $\phi_{\rm gal}$, with $\gamma = -\eth^2 \phi_{\rm gal}/2$. The auto-correlation of the observed galaxy ellipticities is
\begin{equation}
C_l^{\epsilon\epsilon} = - \frac{1}{2}\sqrt{\frac{(l+2)!}{(l-2)!}} \left(C_l^{\gamma \phi_{\rm gal}} +2 C_l^{\epsilon^{\rm s} \phi_{\rm gal}} \right) + C_l^{\epsilon^{\rm s}\epsilon^{\rm s}}
\, ,
\label{eq:epseps}
\end{equation}
where $C_l^{\gamma \phi_{\rm gal}}$ is given by the CMB--shear result, Eq.~\eqref{eq:gammaphi}, with the CMB lensing kernel replaced with the galaxy kernel, i.e.,
\begin{equation}
\frac{\chi_\ast - \chi}{\chi \chi_\ast} \rightarrow W(\chi) \, .
\end{equation}
Similarly, $C_l^{\epsilon^{\rm s} \phi_{\rm gal}}$ is given by the same replacement in the intrinsic--CMB result, Eq.~\eqref{eqn:crosscorr}. Note the additional factor of two in the GI term in Eq.~\eqref{eq:epseps} since both legs of the correlation have a contribution from the intrinsic ellipticities of galaxies. The final term in Eq.~\eqref{eq:epseps} is the II term, first calculated in~\citet{mwk} and~\citet{cnpt}.




The spectra $C_l^{\gamma \phi_{\rm gal}}$ and $C_l^{\epsilon^{\rm s}\phi_{\rm gal}}$ are shown in Fig.~\ref{fig:gi}, for the same galaxy redshift distribution and formation times used in Sec.~\ref{subsec:results}. 
The reduced redshift range of the galaxy window function $W(\chi)$ leads, as expected, to a lower fractional level of contamination from intrinsic alignments (this can be understood as the typical GI correlation increases with increasing redshift separation between the sources;~\citealt{hirataseljak04}). For the redshift distribution considered here, at high multipoles the total GI contamination, $2 C_l^{\epsilon^{\rm s}\phi_{\rm gal}} / C_l^{\gamma \phi_{\rm gal}}$, is of comparable importance to the contamination found for the CMB--galaxy correlation.


\subsection{Effect of galaxy redshift distribution}

\begin{figure}
\centering
\includegraphics[width=0.5\textwidth]{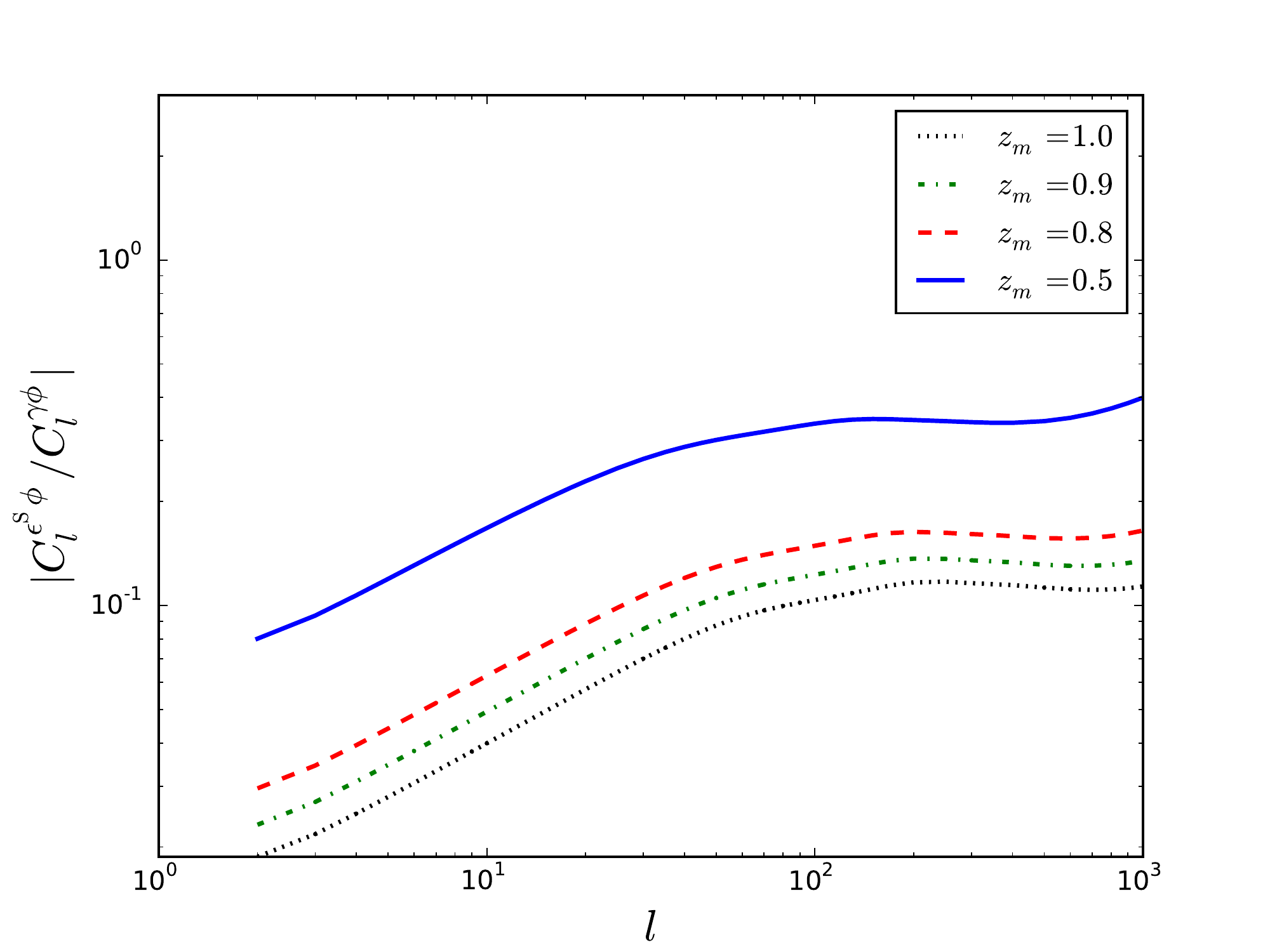}
\noindent
\caption{Absolute value of the ratio $C_l^{\epsilon^{\rm s}\phi}/C_l^{\gamma\phi}$ in the quadratic alignment model (with $z_{\rm P}=3$) for redshift distributions given by Eq.~\eqref{eqn:red} with $z_0=0.5$ (blue solid), $z_0=0.8$ (red dashed), $z_0=0.9$ (green dot-dashed) and $z_0=1.0$ (black dotted).}
\label{fig:red}
\end{figure}
The effect of varying our choice of redshift distribution is considered in Fig.~\ref{fig:red}. 
This shows the ratio of $C_l^{\epsilon^{\rm{s}}\phi}$ to $C_l^{\gamma \phi}$, in the quadratic alignment model with $z_{\rm P} = 3$, for redshift distributions given by
\begin{equation}
f(z) \propto z^2 e^{-(1.4 z/z_0)^{3/2}} \, ,
\label{eqn:red}
\end{equation}
with $z_0= 0.5$, $0.8$, $0.9$ and $1.0$. When $z_0=0.9$, this is the predicted EUCLID redshift distribution from \citet{cosmologyeuclid}.
A strong dependence on the median redshift $z_0$ can be seen (with lower values giving higher contamination fractions, due to the larger galaxy--CMB separation), emphasising the need for accurate redshift distributions in predicting the intrinsic
alignment contamination. 
\altered{In practice there is likely to be a dependence of the time of galaxy formation on the observed redshift distribution, which may alter this effect. }

\subsection{Observational constraints}

There are few observational constraints on intrinsic alignments of blue galaxies at redshifts relevant to cosmic shear surveys. The best constraints to date come from the null detection reported in \citet{wigglez}, which combines
 galaxy shapes from SDSS with spectroscopic redshifts from the WiggleZ Dark Energy survey. Fitting their density--shear correlation to a non-linear alignment model gives 
$95\,\%$ confidence limits of $A= 0.15 ^{+1.03}_{-1.07}$ for a median redshift of approximately $0.6$. Here, $A$ is an amplitude value that scales the alignment amplitude $C$ in Eq.~\eqref{eqn:catelan1} from the fiducial value $C=5\times 10^{-14} h^{-2}\,\rm{M_{\odot}}^{-1}\,\rm{Mpc}^3$ of~\citet{bridleking}.
 The $95\,\%$ upper limit is $A=1.18$, or equivalently a maximum intrinsic alignment contamination to $C_l^{\gamma\phi}$ of approximately $13\,\%$ for the above analysis.
The null detection in \citet{cfhtlens} gives weaker constraints, with a $95\,\%$ upper limit of  $A = 1.84$.


\subsection{Detection forecasts}

In this short section we report the results of a Fisher-matrix analysis to determine the expected constraints on the CMB lensing--galaxy lensing correlation from upcoming cosmic shear data. In particular, combining the shear data from the full Dark Energy Survey (assuming a galaxy number density of $12 \,\rm{arcmin}^{-2}$) with the reconstructed CMB lensing potential from \emph{Planck}~\citep{2015arXiv150201591P}
should give a strong detection of $C_l^{\gamma\phi}$, with a signal-to-noise greater than $20$. This strong detection opens up the possibility of additionally constraining the intrinsic-alignment signal. Jointly constraining an intrinsic-alignment signal, with the shape of the non-linear alignment model of Eq.~\eqref{eqn:linearcrosse} and an amplitude parameter $A$ (with fiducial value $A=1$), alongside cosmological parameters with \emph{Planck} 2015 priors~\citep{2015arXiv150201589P}, would give a $1\,\sigma$ error of $\sigma(A) \approx 0.5$.  An interesting prospect to improve on such a marginal detection is to make use of the expected difference of the intrinsic-alignment amplitudes between samples of red and blue galaxies to null the
cosmic shear signal $C_l^{\gamma \phi}$. This removes much of the cosmic variance from a measurement of the difference of alignment amplitudes. For example, assuming fiducial values of $A_{\rm{red}}=1$ and $A_{\rm{blue}}=0$, and that 20\,\% of the galaxies are red and subject to significant alignments, the difference in amplitudes
$A_{\rm red} - A_{\rm blue}$ could be constrained with a $1\,\sigma$ error of approximately $0.2$.

\section{SUMMARY AND CONCLUSIONS}
\label{sec:summary}

We have considered the impact of intrinsic alignments of spiral galaxies on the CMB lensing--galaxy lensing cross-correlation. Using the quadratic alignment model, we calculated the leading-order cross-correlation between the intrinsic ellipticities of galaxies and the CMB lensing potential, $C_l^{\epsilon^{\rm s}\phi}$, from non-linear evolution. The resulting cross-correlation is very similar in shape to that arising in the linear alignment model, which describes tidal stretching of elliptical galaxies.
An important next step is to attempt to verify these qualitative conclusions with hydrodynamical simulations (e.g.,~\citealt{Codis:2014awa,Chisari:2015qga,illustrus}), for example by comparing the large-scale 3D anisotropic cross-spectrum $P_{\epsilon^{\rm s}_{E}\Phi}(|\vec{k}_\perp|,k_z)$ with our prediction from the quadratic alignment model.

We argued that $C_l^{\epsilon^{\rm s}\phi}$ on large and intermediate scales is dominated by the squeezed limit of the bispectrum of the gravitational potential. This is consistent with the simple physical picture of large-scale modes responsible for the CMB lensing modulating the small-scale power that controls the intrinsic ellipticity through non-linear evolution. We showed by averaging out the small-scale modes in the presence of the large-scale mode why the quadratic alignment model predicts a very similar shape for $C_l^{\epsilon^{\rm s}\phi}$ as the linear alignment model, despite the very different alignment mechanisms at work. The amplitude depends on the level of non-Gaussianity and so has a strong dependence on the time of galaxy formation. Assuming that all intrinsic alignments arise from tidal torquing, we estimated the magnitude of the contamination to the cross-correlation to be of the same order as in the linear alignment model for plausible ranges of the redshift of galaxy formation. 

The similar shapes of $C_l^{\epsilon^{\rm s}\phi}$ in the linear and quadratic alignment models (which extends to the GI term in the case of cosmic shear analyses on large and intermediate scales) justifies the current practice of fitting spiral galaxies with templates derived from the (non-)linear alignment model (e.g., \citealt{wigglez} and \citealt{cfhtlens}). We also note that the assumed difference in amplitude of the intrinsic alignment signals between different galaxy populations could be exploited to reduce significantly the cosmic variance in measurements of the intrinsic alignment signal in upcoming lensing data. 




\section*{ACKNOWLEDGEMENTS}

We thank the participants of the Intrinsic Alignments Workshop held at UCL in July 2015 for useful comments. \altered{We also thank the anonymous reviewer and Alex Hall, Jonathan Blazek and Marcel Schmittfull for comments and questions that improved the manuscript.} PL is supported jointly by the Royal Society of New Zealand--Rutherford Foundation Trust and 
the Cambridge Commonwealth Trust.

\bibliographystyle{mnras}
\bibliography{larsen_updated,challinor}

\appendix
\section{Conditional expectation value of $\epsilon^{\rm s}$ from tidal torquing}
\label{app:condexp}

Here, we calculate the conditional expectation value of $\epsilon^{\rm s}$, from Eq.~\eqref{eq:qeps}, in the presence of a large-scale mode of the linear Lagrangian displacement potential $\psi^{(1)}(\vec{q})$. When further correlated with the large-scale CMB lensing potential, this gives the cross-correlation $C_l^{\epsilon^{\rm s} \phi}$.

\altered{The intrinsic ellipticity depends quadratically on the trace-free tidal tensor $K_{ij}$, so we are led to consider the zero-lag auto-correlation of $K_{ij}$ conditioned on the large-scale mode. Given the shape of the matter power spectrum, the conditional expectation value is dominated by small-scale modes of the tidal tensor. We approximate these using standard results from second-order Lagrangian perturbation theory in an Einstein-de Sitter universe (e.g.,~\citealt{1995A&A...296..575B}). In particular, we use the real-space version of Eq.~\eqref{eq:psi2kspace},
\begin{align}
\psi^{(2)}(\vec{q}) &= \frac{3}{14}\nabla^{-2}_{\vec{q}} \left[\left(\frac{\partial^2 \psi^{(1)}}{\partial q_i \partial q_j}\right)^2 - \left(\nabla^2_{\vec{q}}\psi^{(1)}\right)^2\right] \nonumber \\
&= \frac{3}{14} \left[\frac{1}{2} \left(\partial_i \psi^{(1)}\right)^2 - \nabla^{-2}_{\vec{q}}
\partial_i \left(\nabla^2_{\vec{q}} \psi^{(1)} \partial_i \psi^{(1)}\right)\right] \, ,
\end{align}
and split into small and large-scale modes. Here, $\nabla^{-2}_{\vec{q}}$ is the inverse Laplacian (i.e., free-space Greens function) and, throughout this appendix, partial derivatives with respect to $q_i$ are denoted $\partial_i$.
The relevant split is
\begin{multline}
\psi^{(2)}_{\rm S} = \frac{3}{14} \Bigl[
\partial_i \psi_{\rm L}^{(1)} \partial_i \psi_{\rm S}^{(1)} 
- \nabla_{\vec{q}}^{-2} \partial_i \left(\nabla_{\vec{q}}^2 \psi_{\rm L}^{(1)} \partial_i \psi_{\rm S}^{(1)}\right) \\
- \nabla_{\vec{q}}^{-2} \partial_i \left(\nabla_{\vec{q}}^2 \psi_{\rm S}^{(1)} \partial_i \psi_{\rm L}^{(1)}\right) \Bigr] \, ,
\label{eq:phi2s}
\end{multline}
where the subscripts $L$ and $S$ denote long- and short-wavelength modes of $\psi$. We calculate the conditional expectation value of $\epsilon^{\rm s}$ by using Eq.~\eqref{eq:phi2s} in the correlator
\begin{equation}
\langle K_{{\rm S},ij} K_{\rm S}^{kl} \rangle_{\rm S} = \langle K_{{\rm S},ij}^{(2)} K_{\rm S}^{(1)kl}
+ K_{{\rm S},ij}^{(1)} K_{\rm S}^{(2)kl} \rangle_{\rm S} \, ,
\label{eq:app1}
\end{equation}
where the expectation value is over $\psi^{(1)}_{\rm S}$. Recall that the tidal tensor
\begin{equation}
K_{ij} \equiv D_{ij} \psi = \frac{\partial^2\psi}{\partial q_i \partial q_j}-
\frac{1}{3} \delta_{ij} \nabla_{\vec{q}}^2 \psi \, ,
\end{equation}
and the velocity potential $\psi(\vec{q},t) = \psi^{(1)}(\vec{q}) + 2 a \psi^{(2)}(\vec{q})$ to second order.
We are ignoring the linear-theory contribution to this expectation value \alteredb{(i.e. terms involving $\langle K_{{\rm S},ij}^{(1)} K_{\rm S}^{(1)kl} \rangle_{\rm S}$ in Eq.~\eqref{eq:app1})} since it does not contribute to the correlation with the large-scale CMB lensing potential.
}

\altered{In evaluating Eq.~\eqref{eq:app1}, only terms with an even total number of derivatives acting on short-scale modes of $\psi^{(1)}$ survive the averaging. Considering the first term on the right of Eq.~\eqref{eq:phi2s}, the relevant terms are
\begin{multline}
\langle K_{{\rm S},ij} K_{\rm S}^{kl} \rangle_{\rm S} \supset \Bigl\langle \Bigl(
\partial_i \partial_m \psi^{(1)}_{\rm L} \partial_j \partial^m \psi^{(1)}_{\rm S} +
\partial_j \partial_m \psi^{(1)}_{\rm L} \partial_i \partial^m \psi^{(1)}_{\rm S}  \\
- \frac{2}{3} \delta_{ij} \partial_m \partial_n \psi^{(1)}_{\rm L} \partial^m \partial^n
\psi^{(1)}_{\rm S} \Bigr) D^{kl} \psi^{(1)}_{\rm S} \Bigr\rangle_{\rm S} + (ij)\leftrightarrow(kl) \, .
\end{multline}
These are easily evaluated in Fourier space to give
\begin{equation}
\frac{7}{3a} \langle K_{{\rm S},ij} K_{\rm S}^{kl} \rangle_{\rm S} \supset \frac{8}{15} \langle (\nabla_{\vec{q}}^2 \psi_{\rm S}^{(1)})^2 \rangle \delta_{\langle i}^{\langle k} \partial^{l\rangle} \partial_{j\rangle} \psi_{\rm L}^{(1)} \, ,
\end{equation}
where the angle brackets denote the symmetric, trace-free part on the enclosed indices.
}

\altered{The contribution of the second term on the right of Eq.~\eqref{eq:phi2s} involves
\begin{multline}
\langle D_{ij} \nabla_{\vec{q}}^{-2} \partial_m (\nabla_{\vec{q}}^2 \psi_{\rm L} \partial^m \psi_{\rm S})K^{(1)kl}_{\rm S} \rangle_{\rm S} = - \int \frac{d^3 \vec{k}}{(2\pi)^{3/2}} \frac{d^3 \vec{k}'}{(2\pi)^{3/2}} \\
\bigl[ P_{\psi^{(1)}_{\rm S}}(k) k^{\prime\,2} \psi^{(1)}_{\rm L}(\vec{k}') e^{i \vec{k}'\cdot \vec{q}}
k^m k^{\langle k} k^{l\rangle}\\
\times (k'+k)_m (\widehat{k'+k})_{\langle i} (\widehat{k'+k})_{j\rangle} \bigr] \, ,
\label{eq:app2}
\end{multline}
plus the term from interchanging $(ij)$ and $(kl)$. Here, $\vec{k}'$ is the wavevector of the large-scale mode and $\vec{k}$ of the short-scale modes, so that $k' \ll k$. Expanding to first-order in $k'/k$, we have
\begin{equation}
 (\widehat{k'+k})_{\langle i} (\widehat{k'+k})_{j\rangle} = \hat{k}_{\langle i} \hat{k}_{j\rangle}
 + 2 \frac{k'}{k} \left( \hat{k}_{\langle i} \hat{k}'_{j\rangle} - \hat{\vec{k}}'\!\cdot\! \hat{\vec{k}} \, \hat{k}_{\langle i}\hat{k}_{j \rangle}\right) \, ,
\label{eq:app3}
\end{equation}
and so the leading-order contribution from the second term on the right of Eq.~\eqref{eq:phi2s} evaluates to 
\begin{equation}
\frac{7}{3a} \langle K_{{\rm S},ij} K_{\rm S}^{kl} \rangle_{\rm S} \supset -\frac{4}{15} \langle (\nabla_{\vec{q}}^2 \psi^{(1)}_{\rm S})^2 \rangle \nabla_{\vec{q}}^2 \psi_{\rm L}^{(1)}
\delta_{\langle i}^{\langle k} \delta_{j \rangle}^{l \rangle} \, .
\end{equation}
Terms of this isotropic form do not contribute to $\langle \epsilon^{\rm s}\rangle_{\rm S}$ by symmetry.
}

\altered{The third term on the right of Eq.~\eqref{eq:phi2s} has a similar structure to Eq.~\eqref{eq:app2}, but with $k^{\prime\,2} k^m$ replaced with $k^2 k^{\prime\,m}$ in the Fourier integrals. The effect of this is that the leading-order term in the expansion~\eqref{eq:app3} no longer contributes and the terms that are first order in $k'/k$ must be retained [and additionally in the factor $(k'+k)_m$]; the end result is that the leading-order contribution from the third term on the right of Eq.~\eqref{eq:phi2s} is
\begin{multline}
\frac{7}{3a} \langle K_{{\rm S},ij} K_{\rm S}^{kl} \rangle_{\rm S} \supset -\frac{4}{105} \langle (\nabla_{\vec{q}}^2 \psi^{(1)}_{\rm S})^2\rangle \\
\times \left( 5 \nabla_{\vec{q}}^2 \psi_{\rm L}^{(1)}
\delta_{\langle i}^{\langle k} \delta_{j \rangle}^{l \rangle} + 6 \delta_{\langle i}^{\langle k} \partial^{l\rangle} \partial_{j\rangle} \psi_{\rm L}^{(1)} \right) \, .
\end{multline}
}

\altered{Finally, combining these results, we have at leading-order
\begin{multline}
\frac{7}{3a} \langle K_{{\rm S},ij} K_{\rm S}^{kl} \rangle_{\rm S} = \frac{16}{21} \langle (\nabla_{\vec{q}}^2 \psi^{(1)}_{\rm S})^2\rangle \\
\times \left( 
 \frac{2}{5} \delta_{\langle i}^{\langle k} \partial^{l\rangle} \partial_{j\rangle} \psi_{\rm L}^{(1)} 
- \nabla_{\vec{q}}^2 \psi_{\rm L}^{(1)}
\delta_{\langle i}^{\langle k} \delta_{j \rangle}^{l \rangle} \right) \, .
\end{multline}
The first term on the right does not contribute to $\langle \epsilon^{\rm s} \rangle_{\rm S}$, while for the second term, we use the identity
\begin{equation}
\delta^j_l \delta_{\langle i}^{\langle k} \partial^{l\rangle} \partial_{j\rangle} \psi_{\rm L}^{(1)} =
\frac{7}{12} \partial_i \partial^k \psi_{\rm L}^{(1)} + \frac{13}{36} \delta_i^k \nabla^2 \psi_{\rm L}^{(1)} \, ,
\end{equation}
so that Eq.~\eqref{eq:qeps} gives the complex ellipticity as
\begin{equation}
\langle \epsilon^{\rm s} \rangle_{\rm S} = \frac{8 C}{525} a(\chi_{\rm P}) \langle (\nabla_{\vec{q}}^2 \psi^{(1)}_{\rm S})^2\rangle 
\left(\frac{\partial}{\partial q_x} + i \frac{\partial}{\partial q_y}\right)^2 \psi_{\rm L}^{(1)}
\end{equation}
for the line-of-sight along the $z$-direction. We can express this in terms of the large-scale Eulerian gravitational potential using the leading-order result that
\begin{equation}
\Phi^{(1)}_{\rm L}(\vec{x},t_{\rm P}) = -\left.\left(\frac{3 a^3 H^2}{2}\right)\right|_{t_{\rm P}}
\left. \psi_{\rm L}^{(1)} \right|_{\vec{q}=\vec{x}} \, .
\end{equation}
For an arbitrary direction on the spherical sky, we have
\begin{equation}
\langle\epsilon^{\rm s}(\hat{\vec{n}};\chi)\rangle_{\rm S} =- \frac{8 C}{525\chi^2} 
\left. \left(\frac{2}{3a^2 H^2}\right)\right|_{t_{\rm P}}
\langle (\nabla_{\vec{q}}^2 \psi^{(1)}_{\rm S})^2\rangle 
\eth^2 \Phi_{\rm L}^{(1)} \, ,
\end{equation}
where $\Phi^{(1)}_{\rm L}$ is evaluated at position $\chi \hat{\vec{n}}$ at the formation redshift.
}

\bsp	
\label{lastpage}
\end{document}